\documentclass[aps,prl,twocolumn,showpacs]{revtex4}
\usepackage{graphicx}
\begin{document}
\title{Light Gluino, Light Bottom Squark Scenario, and LEP Predictions}
\author{Kingman Cheung$^1$ and Wai-Yee Keung$^{1,2}$}
\email[Email:]{cheung@phys.cts.nthu.edu.tw,keung@uic.edu}
\affiliation{
$^1$
National Center for Theoretical Sciences, National Tsing Hua 
University, Hsinchu, Taiwan, R.O.C. \\
$^2$
Department of Physics, University of Illinois at Chicago, Chicago, IL 60187
}
\date{\today}

\begin{abstract}
The scenario of light gluinos and light sbottoms 
was advocated to explain the discrepancy between the 
measured and theoretical production of $b$ quarks at the Tevatron.  
This scenario will have model-independent predictions for $Z\to 
q\bar q \tilde{g} \tilde{g}$ at the $Z^0$-pole, and $e^+ e^- \to
q\bar q \tilde{g} \tilde{g}$ at LEPII.  We show that the data for
$Z\to q \bar q g^* \to q\bar q b \bar b$ at LEPI cannot constrain the scenario,
because the ratio $\Gamma(Z \to q\bar q \tilde{g} \tilde{g})/ \Gamma(
Z \to q \bar q g^* \to q\bar q b \bar b) = 0.15 - 0.04$ for $m_{\tilde{g}}=
12-16$ GeV is smaller than the uncertainty of the data.
However, at LEPII the ratio
$\sigma(e^+ e^- \to q\bar q \tilde{g} \tilde{g})/ \sigma(
e^+ e^- \to q \bar q g^* \to q\bar q b \bar b) \simeq 0.4 - 0.2$ for 
$m_{\tilde{g}}= 12-16$ GeV, which may give an observable excess 
in $q\bar q b\bar b$ events;  especially, the $4b$ events.
\end{abstract}
\preprint{NSC-NCTS-020531}
\maketitle

Weak-scale supersymmetry is the leading candidate for physics beyond the 
standard model (SM). Supersymmetry (SUSY) is built on a solid theoretical
and mathematical foundation.  It is also well-motivated as an elegant 
solution to the gauge hierarchy problem and has merits of gauge-coupling
unification, dynamical electroweak-symmetry breaking, and providing a
legitimate candidate for dark matter.  The search for SUSY will be a major
goal at future collider experiments, and in
precision measurements, such as $g-2$ and electric dipole moments 
\cite{report}.

One of the long-standing problems in heavy flavors is the excess in
hadronic production of $b$ quarks recorded by both 
Collider Detector at Fermilab (CDF) and D\O\ Collaborations
\cite{cdf-d0}.  The data is about a factor of 2 larger than the prediction
by the most optimal choice of parameters in perturbative QCD (here
optimal means that the parameters such as $b$-quark mass $m_b$,
the factorization scale $\mu$ have been tuned to maximize the prediction)
\cite{foot1}.
Such a discrepancy was recently interpreted by Berger et al. \cite{berger}
in the scenario of light gluinos and light sbottoms. 
Light gluinos of mass between $12-16$ GeV are pair-produced by $q\bar q$ 
and $gg$ fusion processes.  These are QCD processes and the
cross sections are similar to $b$-quark production.  The gluinos
undergo subsequent decays $\tilde{g} \to b \tilde{b}^*_1 \,/ \, \bar b
\tilde{b}_1$, where the sbottom has a mass $2-5.5$ GeV.  Therefore,
in the final state there are $b \bar b + \tilde{b}_1\tilde{b}^*_1$, in which
the sbottom either remains stable or 
decays into other light hadrons (e.g. via $R$-parity violating
couplings) and goes into the 
$b$-jet.  Thus, gluino-pair production gives rise to inclusive $b$-quark
cross section.   
The mass ranges are chosen so as to
reproduce both the total cross section and the transverse momentum spectrum
of the $b$-quark.
Before Berger et al.'s work, there have been some studies in the light sbottom
and/or light gluino scenario \cite{new}.  However, such a scenario cannot be
ruled out, unless there exists a sneutrino of at most 1--2 GeV.

A light gluino can be established in some
moduli-dominated SUSY-breaking models, and can even be 
the LSP \cite{model}.  The gluino-LSP scenario was studied in 
Ref. \cite{me} (the gluino-NLSP scenario was studied in Ref. \cite{raby}.)
The light-gluino scenario is consistent with cosmological constraints
and does not affect the precision data as long as the squarks are heavy.
However, the implication would be very different if both the gluino and
sbottom are light.
Therefore, the first impression to Berger et al.'s scenario would be that
the scenario easily contradicts other experiments, especially
the $Z^0$-pole data
because of the light sbottom, as well as the collider search for 
light gluinos.  

Berger et al. \cite{berger} can defend their scenario
by arguing that 
(i) all previous light gluino limits are not applicable because either 
the mass range is different or the decay channel of the gluino is different, 
and (ii) the mixing angle of $\tilde{b}_L$ and $\tilde{b}_R$ can be tuned to
a value such that the tree-level coupling of $\tilde{b}_1$ to $Z$ is 
negligible so as not to upset the $Z$ observables.
However, Cao et al. \cite{cao} showed that such a light gluino and a light
sbottom will contribute significantly to $R_b$ via one-loop gluino-sbottom
diagrams.  In order to suppress such contributions, 
the second $\tilde{b}_2$ has to be lighter than about 125 GeV 
(at $2\sigma$ level) in order to cancel the contribution of 
$\tilde{b}_1$ in the gluino-sbottom loop.
Cho \cite{cho} extended the analysis to the whole set of electroweak
precision data and took into account the stop contributions because of
the SU(2)$_L$ symmetry.  He found a similar conclusion that the $\tilde{b}_2$
must be lighter than about 180 GeV at $5\sigma$ level and the left-right
mixing of the stop must be sufficiently large.
On the other hand, Baek \cite{baek} showed that such constraints can be
relaxed if CP-violating phases are allowed in the model.

The light gluino and light sbottom scenario will certainly give rise to 
other interesting signals, e.g., decay of $\chi_b$ into the light sbottom
\cite{lee}, enhancement of $t\bar t b\bar b$ production at hadron 
colliders \cite{rain}, decay of $\Upsilon$ into a pair of light sbottoms 
\cite{cla}, and 
flavor-changing effects in radiative decays of $B$ mesons \cite{becher}.
As mentioned by Berger et al. \cite{berger}, a light-gluino analysis 
was done by Baer, Cheung and Gunion \cite{me}, in which the gluino is assumed
the LSP.  Here in this work we modify the analysis by
letting the light gluino decay into $b$ and $\tilde{b}_1$,
and study the possible constraint and implication at LEP.

In this Letter, we calculate the associated production of a gluino-pair
with a $q\bar q$ pair and compare to the SM prediction of $q\bar q b\bar b$
at both LEPI and LEPII (here $q$ refers to the sum over $u,d,s,c,b$ and we use
the massless quark approximation).
  We show that the current data from LEPI are not
precise enough to constrain Berger et al.'s scenario.  On the other hand,
at LEPII ($\sqrt{s}=189-209$ GeV) the $q\bar q \tilde{g} \tilde{g}$ production
cross section is about $40-20\,\%$ of the SM production of $q\bar q b\bar b$,
which may be large enough to produce an observable 
excess in $q\bar q b\bar b$ events.
Similar conclusions can also be drawn on the $4b$ production.  
Such results are model-independent.  If Berger et al.'s scenario is correct,
the above prediction is unavoidable.  We, therefore,
urge our experimental colleagues at LEP to analyze the $q\bar q b\bar b$
and $4b$ channels.

\begin{figure}[!bh]
\includegraphics[width=3.3in]{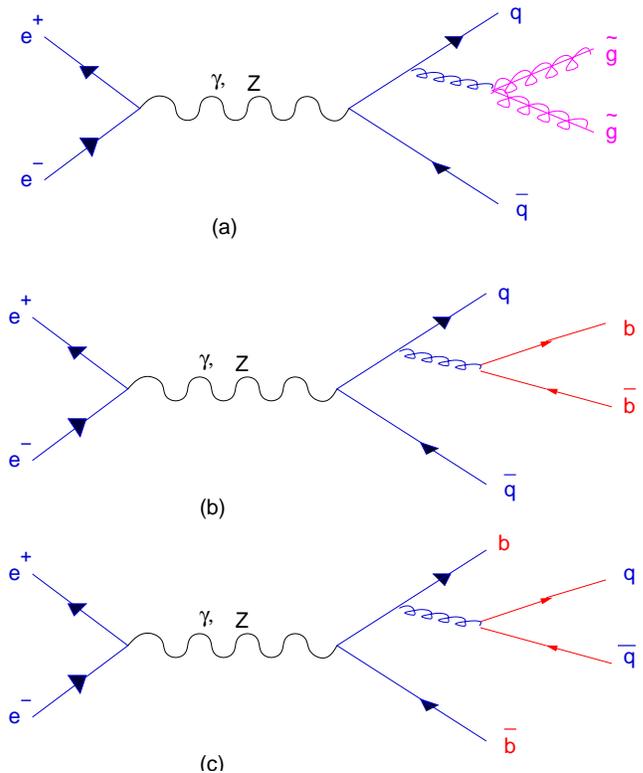}
\caption{\small \label{fig1}
Feynman diagrams contributing to 
(a) $e^+ e^- (Z) \to q\bar q \tilde{g} \tilde{g}$,
(b) $e^+ e^- (Z) \to q\bar q g^* \to q \bar q b \bar b$, and 
(c) $e^+ e^- (Z) \to b\bar b g^* \to b \bar b q \bar q $.
The diagrams with the gluon bremstrahlung off the $\bar q$ are not shown.
}
\end{figure}

At the $Z^0$-pole, 
the lowest-order model-independent channel to produce a gluino-pair is via
a gluon-splitting coming off a quark or anti-quark, as shown in 
Fig. \ref{fig1}(a).  
It is followed by the subsequent decay of gluino $\tilde{g}
\to b \tilde{b}^*_1/ \bar b \tilde{b}_1$, and therefore, it will give rise 
to $q\bar q b\bar b$ production.  The LEP Collaborations had measured
a gluon-splitting process $Z\to q\bar q g^* \to q \bar q b \bar b$ at the
$Z^0$-pole \cite{aleph,opal,delphi}.  The data are given as 
\begin{widetext}
\[
\frac{\Gamma( Z\to q\bar q g^* \to q \bar q b \bar b)}
     {\Gamma( Z\to \; {\rm hadrons})} 
= \left \{
\begin{array}{lr}
(2.77 \pm 0.42 \pm 0.57) \times 10^{-3} & \mbox{ALEPH} \\
(3.07 \pm 0.53 \pm 0.97) \times 10^{-3} & \mbox{OPAL} \\
(3.3 \pm 1.0 \pm 0.8) \times 10^{-3} &    \mbox{DELPHI I} \\
(2.1 \pm 1.1 \pm 0.9) \times 10^{-3} &    \mbox{DELPHI II}
\end{array}
\right . \;.
\]
\end{widetext}
The above data have been corrected for acceptance and cut efficiencies by
each experiment.
We combine the above data assuming that the errors are gaussian, 
each data has equal weights, and the data are uncorrelated.  We obtain the
average and the $1\sigma$ error as 
\begin{equation}
\label{data}
\frac{\Gamma( Z\to q\bar q g^* \to q \bar q b \bar b)}
     {\Gamma( Z\to \; {\rm hadrons})} = 
(2.83 \pm 0.51 ) \times 10^{-3} \;.
\end{equation}

The Feynman diagrams that contribute to the gluon-splitting production of
$Z\to q\bar q g^* \to q\bar q b \bar b$ are shown in Fig. \ref{fig1}(b). 
The Feynman diagrams in Fig. \ref{fig1}(c) contribute to the same final state
but can be easily separated from those in (b) by an invariant mass cut on 
$m_{q\bar q}$.  
In the calculation, we have chosen
$m_b=4.25$ GeV and the strong running coupling is evaluated at $Q^2=
m^2_{b\bar b}$, which is the offshellness in the virtual gluon.  We obtain
in the SM
\begin{equation}
\label{sm}
\left. \frac{\Gamma( Z\to q\bar q g^* \to q \bar q b \bar b)}
     {\Gamma( Z\to \; {\rm hadrons})}  \right |_{\rm SM} = 
2.81 \times 10^{-3} \;,
\end{equation}
where we take the total hadronic width of the $Z$, $\Gamma_{\rm had}=1.745$
GeV \cite{lep-wg}. It agrees well with the data in Eq. (\ref{data}).

Now we proceed to calculate $Z\to q\bar q \tilde{g} \tilde{g}$ to see if
it would contribute at a level larger than the uncertainty 
of the data.   However, we found that
\begin{equation}
\label{gg}
 \frac{\Gamma( Z\to q\bar q \tilde{g} \tilde{g} )}
     {\Gamma( Z\to \; {\rm hadrons})}   =
(0.43 - 0.12 ) \times 10^{-3}  
\;
\end{equation}
for $m_{\tilde{g}} = 12-16$ GeV.
We have chosen $\alpha_s(Q^2 = m^2_{\tilde{g} \tilde{g}})$ analogous
to the $b\bar b$ calculation above.  
It implies that $\Gamma( Z\to q\bar q \tilde{g} \tilde{g} )$ is only
a small fraction (15\% -- 4\% for $m_{\tilde{g}}=12-16$ GeV) of 
$\Gamma( Z\to q\bar q g^* \to q \bar q b \bar b)$, plus
it is less than the $1\sigma$ uncertainty. 
We conclude that the present LEPI data cannot constrain the scenario.
This gluino-pair production is independent of any mixing parameters.

The DELPHI Collaboration \cite{delphi} also measured the $4b$ production 
due to the gluon-slitting. The statistics is even lower.  We would expect 
$Z\to b \bar b \tilde{g} \tilde{g}$ to be subdominant, very similar to
the $q\bar q \tilde{g} \tilde{g}$ case.  We do not pursue it further.

\begin{figure}[bh!]
\includegraphics[width=3.3in]{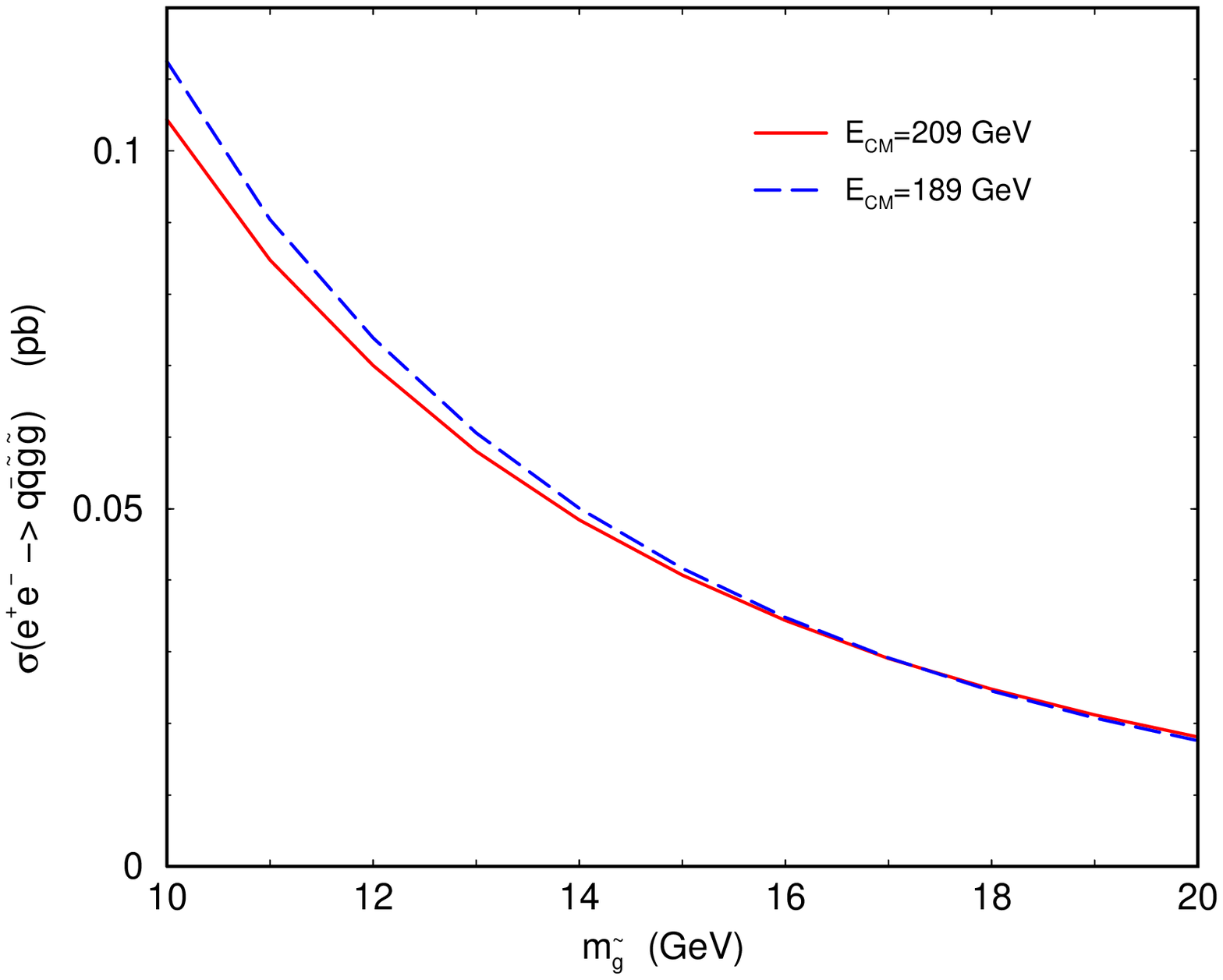}
\includegraphics[width=3.3in]{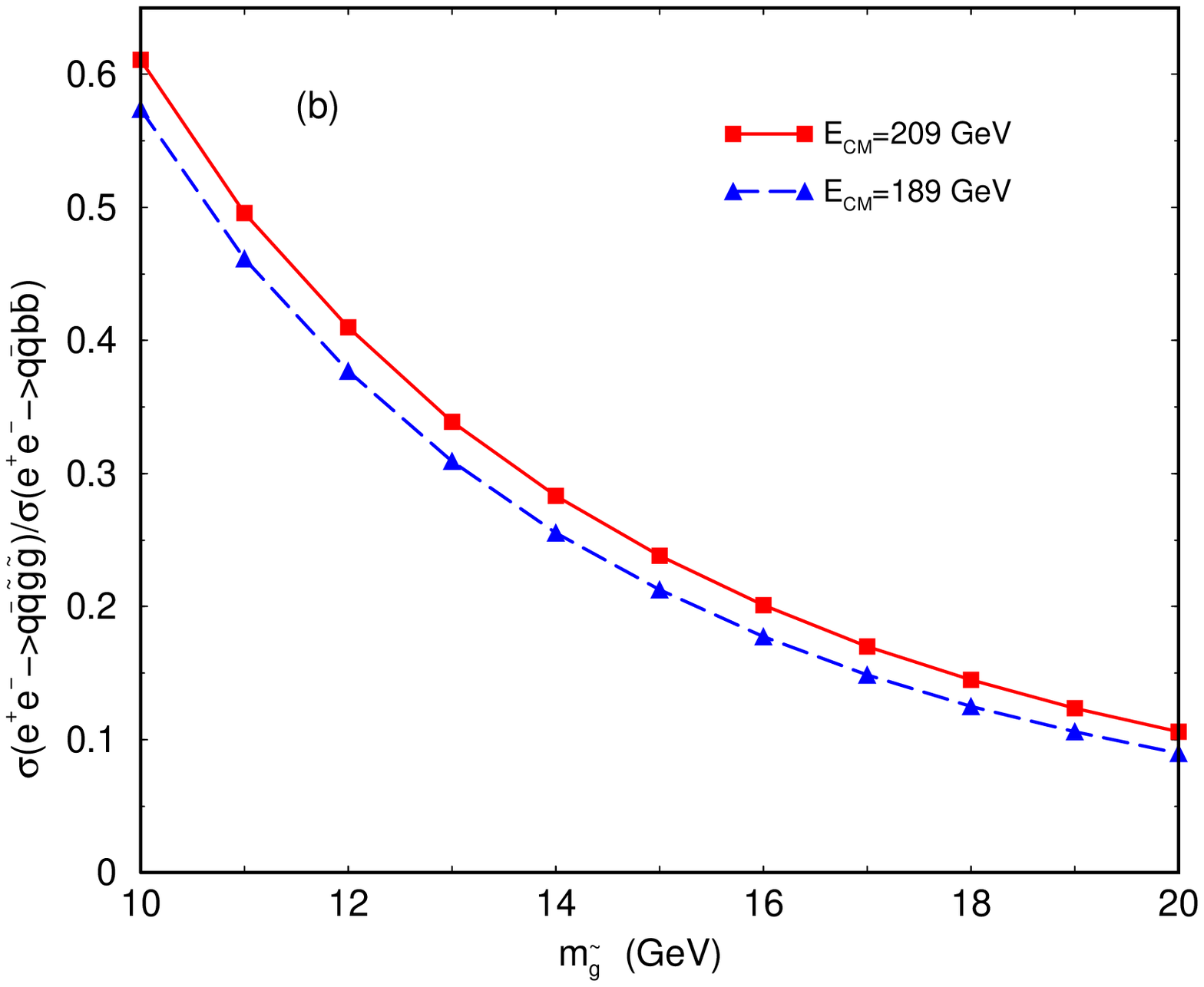}
\caption{\small \label{fig2}
(a) The cross section of $\sigma( e^+ e^- \to q\bar q \tilde{g} \tilde{g})$
versus the gluino mass at $\sqrt{s}=189,209$ GeV.
(b) The ratio $R_{\tilde{g}}$ versus 
$m_{\tilde{g}}$ for $\sqrt{s}=189,209$ GeV.
}
\end{figure}

At LEPII, the situation would be different because of higher energies
and more phase space.  We show the cross section of
$\sigma(e^+ e^- \to q\bar q \tilde{g} \tilde{g})$
versus $m_{\tilde{g}}=10-20$ GeV for $\sqrt{s}=189,209$ GeV in 
Fig. \ref{fig2}(a).  In general, there are two factors affecting the
cross section: (i) this is a $s$-channel process as far as the initial
$e^+ e^-$ is concerned, and so the cross section decreases with $\sqrt{s}$, 
and (ii) as $\sqrt{s}$ 
increases more phase space is available for the massive gluinos.
The cross sections for $\sigma(e^+ e^- \to q\bar q g^* \to q\bar q b\bar b)$ 
with $m_b=4.25$ GeV
are $0.19$ and $0.17$ pb at $\sqrt{s}=189$ and $209$ GeV, respectively. 

In Fig. \ref{fig2}(b), we plot the ratio 
\[
R_{\tilde{g}} \equiv
\frac{\sigma(e^+ e^- \to q\bar q \tilde{g} \tilde{g})}
{\sigma(e^+ e^- \to q\bar q g^* \to q\bar q b\bar b)}
\]
for $m_{\tilde{g}}=10-20$ GeV.  
For the mass range of interest, $m_{\tilde{g}}=12-16$ GeV, the ratio
at $\sqrt{s}= 189\,(209)$ GeV is
\begin{equation}
\label{r}
R_{\tilde{g}} = 
\left \{ \begin{array}{ll}
0.38\, (0.41)& \quad \mbox{for $m_{\tilde{g}}=12$ GeV} \\
0.26\, (0.28)& \quad \mbox{for $m_{\tilde{g}}=14$ GeV} \\
0.18\, (0.20)& \quad \mbox{for $m_{\tilde{g}}=16$ GeV} \;.
\end{array}
\right .  
\end{equation}
Since the rate for gluino-pair production is about $40-20$\% of the 
SM prediction, we would expect an observable 
excess in $q\bar q b\bar b$ events at
LEPII.  We note that the ratio for 
$\sigma(e^+ e^- \to b\bar b \tilde{g} \tilde{g})/
\sigma(e^+ e^- \to b\bar b g^* \to b\bar b b\bar b)$ is very similar.  Though
the $4b$ final state would be more spectacular, the statistics would be a few
times lower.

\begin{figure}[bh!]
\includegraphics[width=3.3in]{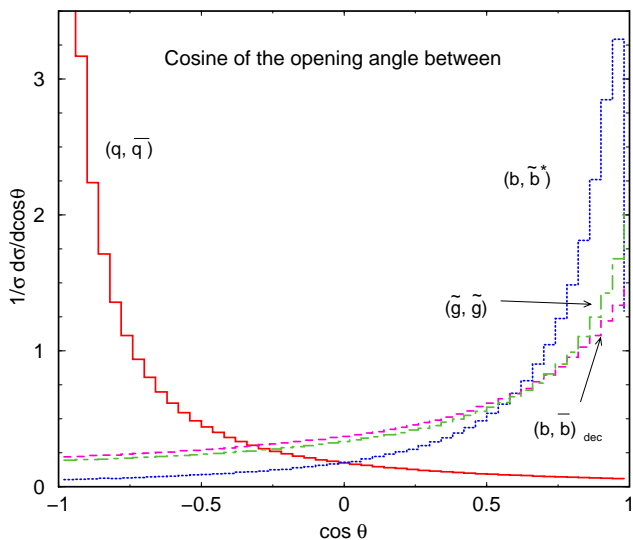}
\caption{\small \label{cos-open}
Distributions of the cosine of the opening angle between the $q\bar q$ pair,
between the decay products, a $b$-quark and a sbottom from a gluino,
between the gluino pair before they decay, and between the two $b$ quarks
from the gluino decay, at $\sqrt{s}=189$ GeV for $m_{\tilde{g}}=12$ GeV.
}
\end{figure}

%%%%%%>> Add ther new figure HERE.
In Fig. \ref{cos-open}, we show the angular separation among the final 
state particles.
The decay products of each gluino, i.e., a $b$-quark and a sbottom, are
very close to each other with $\cos\theta$ peak at above $0.9$.
Experimentally it may be very difficult to separate them.  
Thus, the sbottom will simply go almost along with the
$b$-quark.  The final state then looks like a $q \bar q b \bar b$.
In Fig. \ref{cos-open}, 
we also show the cosine of the 
opening angle between the $q \bar q$ pair, between
the gluino pair before they decay, and between the $b$-quarks decaying
from the gluinos.
The $q\bar q$ pair is back-to-back while the $b$-quarks are very close to
each other.  In addition, the $q$ and  $\bar q$ are very energetic while 
the two $b$'s are soft.  This event topology
is very similar to that of the SM gluon-splitting process.
Thus, we expect the selection efficiencies of the SM 
gluon-splitting process and the
gluino-pair production are very similar.  

 So far, throughout the analysis we used a value $m_b=4.25$ GeV, somewhat
lower than the value employed in Refs. \cite{cdf-d0,berger}.
  The main reason is to make the SM prediction in Eq. (\ref{sm}) close enough
to the $Z^0$-pole data in Eq. (\ref{data}).  If we used $m_b=4.75$ GeV,
the SM prediction would be lower but still within $1.2\sigma$ of the 
data in Eq. (\ref{data}).  
Therefore, the data in Eq. (\ref{data}) could not indicate any excess at a
significant level.  On the other hand, if we change $m_b=4.75$ GeV in the 
LEPII calculation, the results change slightly, giving a slightly 
larger ratio $R_{\tilde{g}}$ of Eq. (\ref{r}):
\begin{equation}
\label{r2}
R_{\tilde{g}} = 
\left \{ \begin{array}{ll}
0.45\, (0.49)& \quad \mbox{for $m_{\tilde{g}}=12$ GeV} \\
0.31\, (0.34)& \quad \mbox{for $m_{\tilde{g}}=14$ GeV} \\
0.21\, (0.24)& \quad \mbox{for $m_{\tilde{g}}=16$ GeV} 
\end{array}
\right .  
\end{equation}
at $\sqrt{s}=189 \,(209)$ GeV for $m_b=4.75$ GeV.  
The observability of excess in $q\bar q b\bar b$ events increases.  
The result in Eq. (\ref{r}) would then be more conservative.

There is another process similar to the one shown in Fig. \ref{fig1}(b) 
with $q,\bar q$ replaced by $\tilde{b}_1, \tilde{b}_1^*$.  However, 
$\tilde{b}_1$ couples to the photon with an electric charge $-1/3$ but not 
to the $Z$ in Berger et al.'s scenario.  Furthermore, it is a scalar.  We,
therefore, expect this process to be sub-dominant to the one that we 
are considering here.  Nevertheless, it gives an additional, yet small,
contribution to the excess in $q\bar q b\bar b$ events.

The effect of including the light gluino and sbottom into the running of
the strong coupling constant is rather mild \cite{chiang}.  
The difference in $\alpha_s$
is only 6\% (3\%) when we run the scale down from $M_Z$ to $24$ GeV $(M_Z/2)$.
Thus, this will not affect our result significantly.  

Each LEP experiment recorded more than $600\,{\rm pb}^{-1}$ luminosity
for energy between 183 and 209 GeV, with most luminosity at 189 and 207 GeV
\cite{lep-wg}.  With a total luminosity more than $2\,{\rm fb}^{-1}$ 
collected by four
experiments, there should be sufficient number of $q\bar q \tilde{g} \tilde{g}$
signal events above the gluon-splitting background.  However, 
at energies above 
$2M_W$ other backgrounds such as $WW,ZZ \to 4\,{\rm jets}$ have to be
discriminated also.  Since the $q\bar q$ pair is back-to-back and energetic
while $\tilde{g}\tilde{g}$ or $b\bar b$ pair tends to soft
and rather close to each other, one 
can make use of this event topology to discriminate the signal from 
the 4-jet events of $WW$ or $ZZ$ decays.  Contamination from gluon-splitting
into other light quarks can be reduced by displaced vertices.  
Detailed detector-dependent analysis is beyond the scope of the present
paper.  

After selective cuts to reduce backgrounds, the
number of gluon-splitting $e^+ e^- \to q\bar q g^* \to q\bar q b\bar b$
events can be counted.  If an excess in such events is observed, it may 
be due to gluino-pair production followed by the gluino decay 
$\tilde{g} \to b \tilde{b}^*_1 / \bar b \tilde{b}_1$ 
that is discussed in the present paper \cite{foot2}. 
Such a scenario of light gluinos and light sbottoms is advocated 
by Berger et al. to explain the excess in $b$-quark production at
the Tevatron.  
In Fig. \ref{fig2} and in Eq. (\ref{r}), we have shown that the gluino-pair
production is a significant fraction of the production of
$q\bar q b\bar b$ by gluon-splitting.  In principle, it should be observed
if the light gluino and light sbottom scenario is correct.  This 
prediction is independent of the light sbottom coupling to $Z$ boson,
the mass of the second $\tilde{b}_2$,  or
the $\tilde{b}_L - \tilde{b}_R$ mixing angle.

In this Letter, we have calculated the associated production of a gluino-pair
with a $q\bar q$ pair and compared to the SM prediction of $q\bar q b\bar b$
at both LEPI and LEPII.
 We have shown that the current data from LEPI are not
precise enough to constrain Berger et al.'s scenario.  On the other hand,
at LEPII the $q\bar q \tilde{g} \tilde{g}$ production
is about $40-20\,\%$ of the SM production of $q\bar q b\bar b$
by gluon-splitting, 
which may be large enough to produce an observable excess in 
$q\bar q b\bar b$ events.
A similar conclusion can also be drawn on the $4b$ production.  
If Berger et al.'s scenario is correct, the prediction here is
unavoidable.  We, therefore,
urge our experimental colleagues at LEP to analyze the gluon-splitting
$q\bar q b\bar b$ and $4b$ events.
Wishfully, this is a sign of supersymmetry.

We would like to thank Gi-Chol Cho for enlightening discussions on the 
subject of electroweak precisions, and to Alex Kagan for an interesting
discussion.
This research was supported in part by the NCTS under a grant from the 
NSC of Taiwan Republic of China.

%%%%%%%%%%%%%%%%%%%%%%%%%%%%%%%%%%%%%%%%%%%%%%%%%%%%%%%%%%%%

%%%%%%%%%%%%%%%%%%%%%%%%%%%%%%%%%%%%%%%%%


\begin{thebibliography}{99}
\bibitem{foot1}
References \cite{cacc} argued that if the most up-to-date $B$ 
fragmentation function is used the observed excess can be reduced to
an acceptable level.  Fields \cite{field} also pointed out
the contributions from fragmentation of gluon and light quarks.
\bibitem{foot2}
Because of the Majorana nature of gluinos, there could be 
$bb$, $\bar b \bar b$,
or $b\bar b$ in the final state.  Given that the cross section is of order
0.05 pb, the event rate may not be large enough to have a reliable count of
such final states unless the $b$-tagging efficiency is high.
\end{thebibliography}
\end{document}